# Ionospheric (H-atom) Tomography: a Feasibility Study using GNSS Reflections.

GIOS-1
ESA Contract No. Starlab/CO/0001/02

## Abridged Final Report

## Feasibility Study

**ESA/ESTEC Technical Manager: B. Arbesser-Rastburg**

Prepared by
Josep Marco, Giulio Ruffini and Leonardo Ruffini
Approved by Giulio Ruffini
giulio@starlab-bcn.com
*Starlab, December 2002,*
*Edifici de l'Observatori Fabra, C. de l'Observatori s/n,*
*Muntanya del Tibidabo, 08035 Barcelona, Spain*
http://starlab.es

## ABSTRACT

In this report we analyze the feasibility of ionospheric monitoring using GNSS technology. The focus will be on the use of LEO GNSS data, exploiting GNSS Reflections, Navigation and Occultation TEC measurements.

In order to attack this question, we have simulated GNSS ionospheric TEC data as it would be measured from a polar LEO (exploiting Navigation, Occultation and Reflection TEC data) and IGS ground stations, through the use of a climatic ionospheric model (we have explored both NeQuick and PIM).

We have also developed a new tomographic approach inspired on the physics of the hydrogen atom, which has been compared to previous successful but somewhat awkward methods (using a voxel representation) and employed to retrieve the Electronic Density field from the simulated TEC data.

These tomographic inversion results using simulated data demonstrate the significant impact of GNSS-R and GNSS-NO data: 3D ionospheric Electron Density fields are retrieved over the oceans quite accurately, even as, in the spirit of this initial study, the simulation and inversion approaches avoided intensive computation and sophisticated algorithmic elements (spatio-temporal smoothing).

We conclude that GNSS-R data can contribute significantly to the GIOS (Global/GNSS Ionospheric Observation System).

# 1. INTRODUCTION

Ionospheric Electron Content measurements are an important element for Space Weather research and operations. Adverse conditions in the space environment can cause disruption of satellite operations, communications, navigation, and electric power distribution grids, leading to a variety of Socio-economic losses. The initial focus of this program will be on better coverage of data-void or data-sparse regions (e.g., data over the oceans, complementary data). Little data on ionospheric electron content is presently available over the oceans, although this situation will be mitigated by occultation measurements (CHAMP, COSMIC), and the vertical character of GNSS-R soundings together with their availability over water (and perhaps ice and land) covered areas will be able to fill these gaps.

It is well known that the atmosphere affects the propagation of radio signals. Both the neutral troposphere and the ionosphere have an impact on ranging measurements from radar systems. In fact, it has been an important goal for the GPS/GNSS research community to test the limits of the geophysical measurement techniques derived from this technology. Both the troposphere and the ionosphere have been an object of intense research exploiting the fact that GPS (L band) signals are susceptible to the atmospheric gas and plasma distribution. This, tied to the high precision of the GPS system, has opened a wide door to study atmospheric phenomena.

Because of the existence of ionised free electrons, the ionosphere adds a delay of a few meters to the GNSS signal (L band). The exact amount depends on the Electronic Density along the ray link path and on which of the GNSS available frequencies is considered (e.g., in GPS, f1 = 1.57542 GHz and f2=1.22760 GHz). The dispersive nature of this phenomenon is exploited to measure the integrated free electron content delay accurately, and, if needed, to remove it from the measurements (as in dual frequency GPS, for example). Consider a signal travelling at time t, between a given satellite and receiver, and let $I = \int_{ray} \rho(\bar{x}) dl$ be the integrated electron density along the ray traversed on by the signal (in electron per square meter). Then the delay at $D_i$ is modelled by:

$$D_i = D_{s.l.}(t) + I\alpha/f_i^2 + T + \ddot{c}_{sat}^i + \ddot{c}_{rec}^i + \text{noise} + (D^i - D_{s.l.}),$$

where $\alpha$=40.30 m$^3$/s$^2$, $D^i$ is the geometric length of the real ray, $D_{s.l.}$ is the length of the ray if it travelled in a straight line (in the vacuum), T models other frequency independent terms, and $c^i_{sat}$ and $c^i_{rec}$ are the instrumental biases (which are assumed to remain constant). The last term is the difference between the length of the real ray and the length of the ray if it propagated in the vacuum and is also small for non-grazing geometries.

Ionospheric delay on GPS depends on the **Integrated Electron Density**, or **Total Electron content (TEC)** along the **ray link path** and is usually expressed in electrons per square meter divided by 1e16 (TEC units or TECU). Electron Density is usually expressed in number of free electrons per cubic meter (**ED**). Typical peak value ED are of the order of 1E12 electrons per cubic meter and are found around 300 km of altitude.

An approach for tomography of the ionosphere is through the **voxel representation** [Ruffini, 1998]. An important problem in voxel tomography is that it is in general an ill-determined problem: the data is typically not sufficient to uniquely specify a solution. Many approaches are possible to address this problem. The problem in the present approach is linear, of the form y= A x. The matrix A is the "integration matrix". It has as many rows as there are data measurements (one for each ray), and as many columns as there are unknowns (one for each voxel plus the bias constants). Its entries, then, are the lengths of the ray portions spent in each voxel, plus a 1 or 0 depending on the satellite and receiver involved (in the bias sector of A). The equation y=Ax is converted into a chi-square problem: we have to minimize $(y-Ax)^2$. There are many solutions, however, and this is reflected by the fact that A^T A has zero determinant. The approach we will typically take to solve this problem is to add a constraint, and minimize $(y-Ax)^2+(Bx)^2$. These constraints are normally smoothing constraints. This is a natural choice. If data is missing in some portion of the grid we ask that the solution interpolates x using the data from other places. Thus, the grid does not set the effective resolution of the solution, which can be coarser depending on data availability. It is important to note that the constraint will have strong effect in areas of low data availability, while in areas with abundant data they will not interfere too much. The effective resolution of the system is thus not homogeneous, but will vary depending on the area's data availability—a good point.

A promising approach for tomography of the ionosphere is the **ingestion** of GPS data into models. In previous work [Ruffini 2001] we investigated the possible use of spaceborne GNSS Reflections (GNSS-R) to sound the ionosphere, a possibility already discussed in [Komjathy 1996]. [Ruffini 1998] analysed GPS data to extract information about the ionospheric electron density distribution. [Ruffini 1999] ingested GNSS TEC data into PIM using a simple least-squares approach, and obtained good fits to GPS data (40 cm of post-residual mean, compared to 30 using full blown tomographic methods involving many more parameters). The key for carrying out this idea was the availability of the integration matrix from the tomographic approach, which was used as observation operator in the ingestion functional.

The possibility of using **GNSS-R ionospheric data** is of high interest: GNSS-R would provide a higher number of vertical, or at least oblique, bi-TEC measurements over the oceans, an important missing piece in a future GNSS Ionospheric Monitoring System. It is known that ionospheric electron content data measured along vertical directions, when ingested in global ionospheric models, highly enhances the accuracy of such models since they complement occultation soundings. Little data with these characteristics is presently available over the oceans, and the vertical character of GNSS-R soundings together with their availability over water surfaces (and perhaps ice or land) covered areas will be able to fill these gaps.

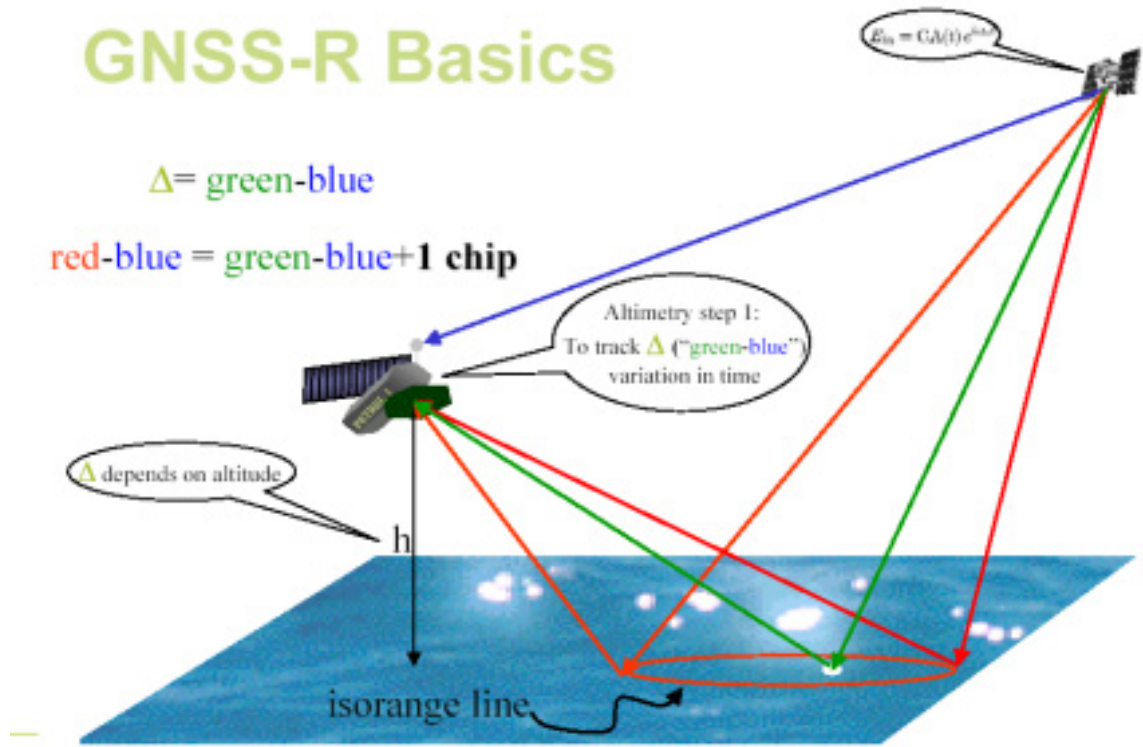

**Figure 1** Basic GNSS-R ranging concept.

As concluded in WP3500 of the Paris Beta ESA/ESTEC project [Ruffini 2002], Ionospheric electron content data estimated from GNSS-R can and should be ingested in Global Ionospheric models.

The accompanying slides to this presentation are available at [Ruffini 2002b]

## 2 SIMULATING GNSS TEC data

Our goal in this feasibility study is to perform tomography of the ionosphere using the combined data of GNSS LEO (navigation, occultation and reflections) and ground data in a synergy, to produce realistic 3D models of the electron density content of the atmosphere. To tackle this problem we have simulated the ionosphere using PIM (a climatic model) and we use a virtual GPS constellation and a single LEO satellite. Simulations have been carried out using the current GPS constellation and a polar LEO orbiter, as well as the current IGS ground network. A polar orbiter is a good choice, as discussed in the Starlab Petrel proposal, as it provides global coverage for many applications.

Note, however, that for the purposes of ionospheric monitoring, the conditions for observation are rather different than for other types of EO. This is because the ionosphere is quasi-static in an inertial frame. Therefore, a single LEO will only be able to sample a rather fixed slice of the ionosphere. Ground stations and filtering techniques can be used to propagate the slice solution elsewhere, but at a price in precision.

Only one LEO orbit has been implemented in our simulation. This is because the goal of this initial study was limited to the demonstration of the impact of GNSS-R data in the GIOS. As discussed, GNSS-R data has the **biggest impact over the oceans**, where there is presently no data, and where GNSS-ON data will only able to provide horizontal TEC. For the purposes of demonstration, we aimed to show that over a given slice over the oceans we could produce a sensible ED solution.

In the simulation we have considered the following 3 sources of ionospheric (slant) TEC data:

- **GNSS-G data:** this TEC data is produced from the present 300+ IGS stations on the ground. These stations provide a steady source of mostly vertical TEC around the globe.
- **GNSS-ON data:** this is TEC data produced from GNSS LEO occultation and navigation: any LEO-to-GPS link TEC. Ray links for GNSS-ON data will typically be colour coded green in the graphs below.
- **GNSS-R data:** this denotes GPS-to-Ground-to-LEO TEC. Ray links for GNSS-R data will be colour coded red.

**Gaussian noise** was also added to the data. Using the above-discussed figures, we added an error of 10 TECU after 10 seconds of averaging (a rather conservative assessment) for GNSS-R data, as we now discuss. The other types of GNSS data are not very affected by noise (sub-TECU).

According to present models and experiments, GNSS-R code bistatic ranging data has an intrinsic noise of less than 0.5 meter (about 5 TECU) after 1 second of integration, with the assumption of a reasonably large antenna (25-28 dB). Thus, this data holds great potential value to ionospheric science, even if only code ranging data is available and the phase cannot be tracked. Ionospheric tomography will benefit greatly from GNSS-R data for an additional reason: as mentioned above, it complements the geometry of occultation data (with very horizontal ray paths) very nicely, at least over the oceans.

Recall that for the GPS $L_I$ combination (using $L_1$ and $L_2$) 1 TECU is equivalent to 10.5 cm of $L_I$ delay and that typical vertical TEC is between 0 and 50 TECU (see next Figure).

GPS and Galileo are designed as multi-frequency systems because the ionospheric contribution to the delay can be removed by making use of its dispersive nature. In fact the "ionosphere-free" linear combination of delays to a good approximation does not depend on ionospheric electron content. The problems of gathering data from a reflection situation and from an occultation are fairly similar. It is also worth mentioning that even if only one frequency is available, comparison of phase and group ionospheric delays (of equal magnitude but opposite sign) can yield ionospheric electron estimates.

We discussed above the potential accuracy from a high gain mission, e.g. with an antenna of 25-28 dB. In the context of low gain missions (with antenna gains of about 15 dB), such dual frequency code measurements would also be of great interest for ionospheric applications. Using dual frequency code pseudo-ranges, current models predict that the ionospheric combination double-slant delays could be measured to better than 2 meter after 1 second of integration, leading to vertical TEC measurements of about 15 TECU accuracy after 1 second, or about 3 TECU after 20 second averaging.

The expected accuracy of single frequency code-phase delay measurements would also be very useful (the error being dominated by the code part of the measurements).

The use of phase measurements would increase even more the interest of ionospheric GNSS-R measurements. We have seen indications in our simulations that the phase of the ionospheric combination is very well behaved after scattering, as expect from the use of an "infinite" synthetic wavelength (using the language in the PARIS Interferometric Processor context)—see [Ruffini 2002].

This could imply that accurate GNSS-R ionospheric phase measurements will be possible from space. Moreover, the behavior of ionospheric phase, according to our simulation, is not sensitive to sea state conditions.

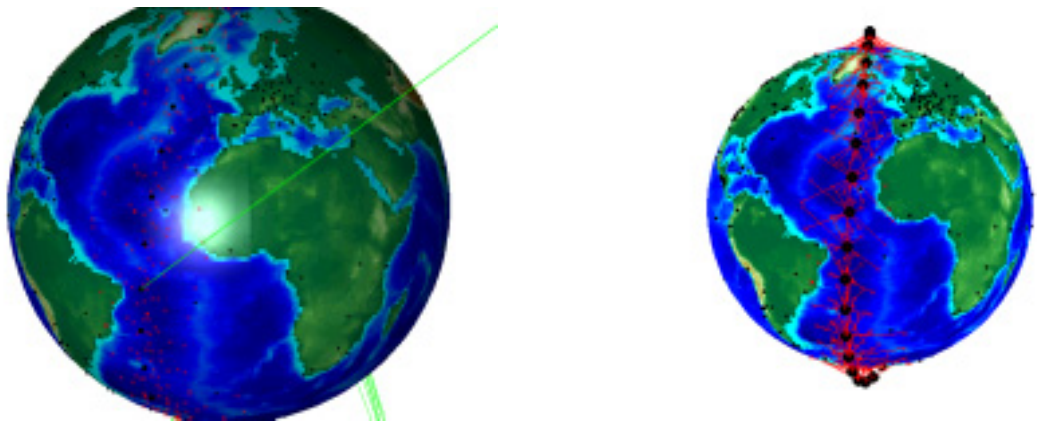

**Figure 2** In the left a grazing occultation ray link giving a particularly large slant TEC (about 600 TECU). Note that reflections can be observed on the surface, and that ground reflections have not been eliminated. In the right LEO positions with a cadence of 3 minutes (orbit positions are shown as black dots). In addition, LEO to Earth Specular links (over the oceans) are shown in red. Ground stations are shown as black dots but no GPS-ground links are shown. Land reflections are identified and neglected (not linked).

For ease of computation, TEC data was only provided with a rather low cadence of 3 minutes. This is a significant factor, as in reality much more data will be available. Nonetheless, we believed (correctly, as it turns out), that even with this meagre data rate we could demonstrate the power of this technique.

We decided to use Parameterised Ionospheric Model (PIM) to generate 3D ED fields with a resolution of about 10 degrees, and developed our own integration routines to produce TEC along any geometry.

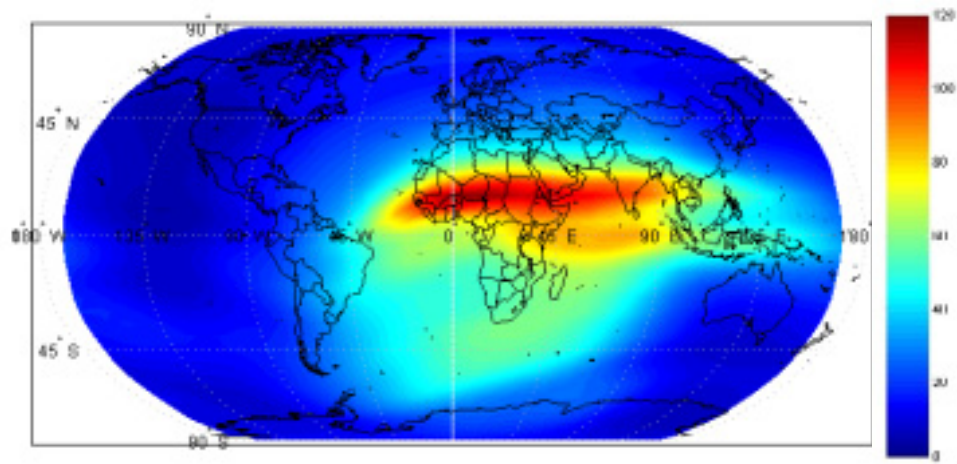

**Figure 3** Simulated ionosphere using Parameterised Ionospheric Model (PIM). Units are TECU. 1992, DOY 270, UTC 1200.

# 3 TOMOGRAPHIC APPROACH: THE H-REPRESENTATION

Previous approaches have focused on voxel representations. Voxels are local support representations. This fact alone makes them inefficient when not the whole ionosphere is sampled---as will be the case here and rather generally.

We have worked with a new representation developed at Starlab (which we call the *H-representation*), based on the solutions to the Schrodinger equation for the Hydrogen atom.

This representation is non-local. This means that if data is available at only specific regions of the ionosphere, all the coefficients in the representation can contribute to the fit. This allows for a good fit where there is data at the expense of sparsely sampled regions. The H-Representation also offers the advantage of easy integration of smoothing terms to account for data scarcity.

Previous approaches have focused on voxel representations, and we are aware of their limitations. Voxels provide a local support representation. This makes them inefficient when not the whole ionosphere is sampled—as will be the case here and rather generally. In addition, they provide a discrete and non-homogeneous (in volume) support representation.

The H-representation is non-local. This means that if data is available at only specific regions of the ionosphere, all the coefficients in the representation can contribute to the fit. This allows for a good fit where there is data at the expense of sparsely sampled regions.

It also offers the advantage of easy integration of smoothing terms to account for data gaps.

The representation used is of the form:

$$ED_{n,l,m}(r,\theta,\phi) = R_{n,l}(r) Y_{lm}(\theta,\phi), \qquad (1)$$

following the Schrödinger solution to Hydrogen atom. Here $Y_{lm}$ are the spherical harmonics, and $R_{n,l}$ is the radial function:

$$R_{n,l}(r) = c_{n,l}\, e^{-\rho/2}\, \rho^{l+1} L^{2l+1}_{n-l-1}(\rho), \qquad (2)$$

where $c_{n,l}$ are normalization coefficients that ensure that:

$$\int_0^{25000} \left( c_{n,l}\, e^{-\rho/2}\, \rho^{l+1} L^{2l+1}_{n-l-1}(\rho) dr \right)^2 = 1. \qquad (3)$$

Similarly, $L_{n-l-1}^{2l+1}$ are the Laguerre polynomials defined by the equation:

$$L_n^k(x) = \sum_{m=0}^{n} (-1)^m \frac{(n+k)!}{(n-m)!(k+m)!m!} x^m, \qquad (4)$$

[Arfken GB and Weber HJ. 1995], and $\rho$ is defined as:

$$\rho = \frac{2r}{na_0}. \qquad (5)$$

In the hydrogen atom, $a_0$ corresponds to the Bohr radius. In our case this is a parameter to determine to optimize the fit. Our simulations show that good fits are given by $a_0$ between 20 and 30. This, however, will depend on the largest value of n allowed (see discussion below).

Note also that in equations (2) we have added an additional $\rho$ in the equation compared with the hydrogen solution for computation purposes.

We want to solve the following system:

$$\text{TEC} = M \cdot x, \qquad (6)$$

where

$$TEC = \gamma \int ED \cdot dl. \qquad (7)$$

Following equation (1):

$$TEC = \int \sum_n \sum_l \sum_m a_{lmn} f(r,\theta,\phi) dl, \qquad (8)$$

which yields:

$$TEC = \sum_n \sum_l \sum_m a_{nlm} \int f(r,\theta,\phi) dl, \qquad (9)$$

so the array x is the array of the coefficients $a_{nlm}$ (our unknowns), and the matrix M is the matrix of integrals, which has the following form:

$$M = \begin{pmatrix} \int R_{10}(r)Y_{00}(\theta,\phi)dl_1 & \int R_{20}(r)Y_{00}(\theta,\phi)dl_1 & \int R_{21}(r)Y_{10}(\theta,\phi)dl_1 & \text{Re}\left(\int R_{21}(r)Y_{11}(\theta,\phi)dl_1\right) & \text{Im}\left(\int R_{22}(r)Y_{11}(\theta,\phi)dl_1\right) & \cdots \\ \int R_{10}(r)Y_{00}(\theta,\phi)dl_2 & \int R_{20}(r)Y_{00}(\theta,\phi)dl_2 & \int R_{21}(r)Y_{10}(\theta,\phi)dl_2 & \text{Re}\left(\int R_{21}(r)Y_{11}(\theta,\phi)dl_2\right) & \text{Im}\left(\int R_{22}(r)Y_{11}(\theta,\phi)dl_2\right) & \cdots \\ \vdots & \vdots & \vdots & \vdots & \vdots & \vdots \\ \int R_{10}(r)Y_{00}(\theta,\phi)dl_n & \int R_{20}(r)Y_{00}(\theta,\phi)dl_n & \int R_{21}(r)Y_{10}(\theta,\phi)dl_n & \text{Re}\left(\int R_{21}(r)Y_{11}(\theta,\phi)dl_n\right) & \text{Im}\left(\int R_{22}(r)Y_{11}(\theta,\phi)dl_n\right) & \cdots \\ \vdots & \vdots & \vdots & \vdots & \vdots & \vdots \end{pmatrix}$$

(10)

while the array x has the form:

$$x = \begin{pmatrix} a_{100} \\ a_{200} \\ a_{210} \\ a_{211R} \\ a_{211I} \\ \vdots \end{pmatrix}. \quad (11)$$

So, given the TEC for each ray link and M, we are to solve

$$x = M^{-1} \cdot TEC, \quad (12)$$

where $M^{-1}$ is the Penrose pseudo-inverse.

The inversion can also be regularized using smoothing terms if a high number of coefficients are desired (we leave this for future work).

The number of unknowns increases following the relation:

$$n = \sum_{i=1}^{n} i^2. \quad (13)$$

We note that other representations are possible and should be investigated.

## 4 TOMOGRAPHIC RESULTS

As can be seen in the Figures below, the H-representation provides an efficient way to represent the solution space.

With as little as n=5 (55 coefficients) we obtained a fit of 7 TECU under the LEO track, using only LEO data (Figure 20), will with the addition of IGS ground data gave a fit of about 13 TECU.

The addition of ground data from a few stations provided a better global fit, as expected.

In Figure 4 we see the results of the n=8 TEC fit.

The main results are in the following table:

| n | # coeff | $a_0$ (km) | $\chi$ TEC (TECU) | $\chi$ Ring ED (Tera el/m$^3$) |
|---|---|---|---|---|
| 5 | 55 | 30 | 15.75 | 0.19/0.19 |
| 6 | 91 | 10 | 11.35 | 0.35/0.45 |
| 6 | 91 | 20 | 10.20 | 0.17/0.20 |
| 8 | 204 | 10 | 7.26 | 0.28 |
| 8[1] | 204 | 10 | 5.33/4.87 | 0.15/0.16 |
| 8 | 204 | 20 | 6.34 | 0.27 |
| 8 | 204 | 10 | 12.22 | 0.18 ($\lambda$=5000) |
| 8 | 204 | 10 | 10.06 | 0.17 ($\lambda$=3000) |
| 8 | 204 | 10 | 8.06 | 0.15 ($\lambda$=1500) |
| 8 | 204 | 10 | 7.29 | **0.15 ($\lambda$=500)** |

**TABLE1**- Results for differents parameters of the H-representation. Xi-errors were computed in the TEC ionosphere (fouth row) and in the ED ring (fifth row). The first result in the fifth row is the Xi with GNSS-R data and the second one is the Xi without GNSS-R data. The last four columns are calculated with a smothing constraint.

We show the H-order and number of coefficients, the $a_0$ parameter, the slant TEC fit, and the ring fit, with/without GNSS-R data.

Thus, we see that as expected, as the order of n and the number of coefficients increase a better fit results. It is clear that using GNSS-R data improves the ED results on the Orbital Ring (see Figure 22 for an explanation of this concept), as expected, although a bigger impact is expected at higher n. Figure 22 and onwards illustrate ED profiling with and without GNSS-R data to highlight the above numbers.

As expected, the $a_0$ parameter must be tuned to avoid putting weight above the area of interest. According to the formula for the Bohr radius, $r_0 = n^2 a_0$. Thus, we should seek to have the support from the radius functions below 1000 km.

In the last four rows (blue) we show preliminary solutions with a smoothing constraint penalizing high n components of the solution (this is done by adding a term of the form $\lambda[n^2]$ to the $\chi^2$).

Finally, recall that spherical harmonic representation resolution is of the order of 180/n. Thus, for n=8 we are at about 22 degrees of spatial horizontal resolution, and about 100 km vertical resolution.

At n=14 we would have about 1000 unknowns, a horizontal resolution of 13 degrees, and about 50 km vertical resolution. At n=20 we have almost 3000 unknowns, for 9 degrees of resolution. This is the resolution we actually used for PIM. To these unknowns we will have to add the emitter-receiver biases (one per GPS satellite and GPS receivers).

---

[1] Note that here we used a 1 minute data cadence—if not, the solutions where degenerate without the addition of constraints. With the 3 minute cadence. No ground data was used, only LEO.

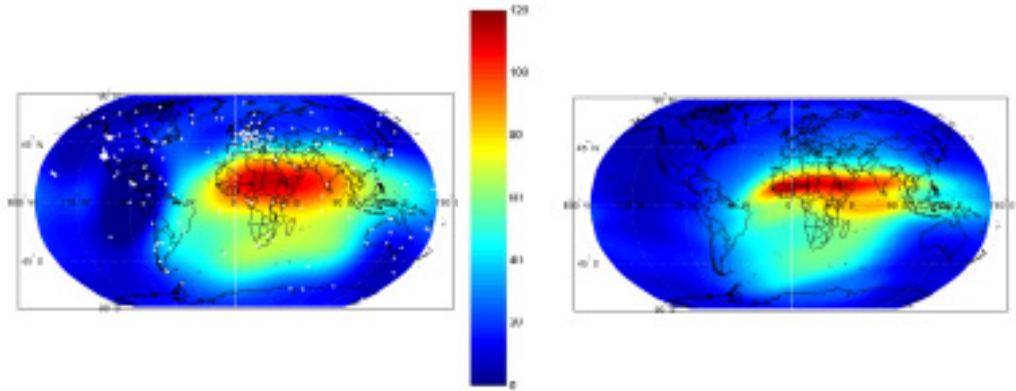

**Figure 4** Top: Recovered TEC with n=8 (204 coefficients, a=20), using ground data (IGS stations appear as white dots) and all LEO data (over 10,000 measurements). Residual variance is of 6.3 (slant) TECU, mainly due to model "quantisation". In the bottom we see the PIM ionosphere.

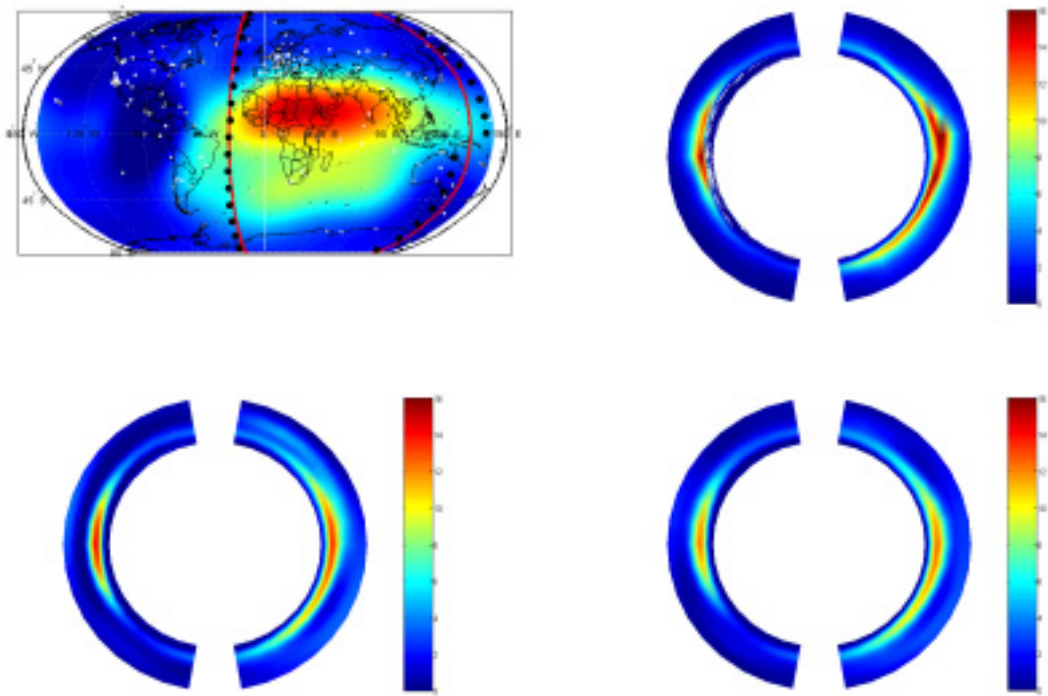

**Figure 5** Solution with an H-representation of order n=8 (204 unknowns, a=10 km) with GNSS-R data (top, 0.15 Tera el/$m^3$ mean error), reference truth (middle) and with a n-constraint (of $\lambda\, n^2$ type) (0.14 Tera el/$m^3$ mean error). No ground data has been used. Altitude spans from 0 to 1,000 km of altitude. Units are ED/1d11 electrons per cubic meter.

# 5 CONCLUSIONS

In this study a new representation has been used in order to model the ionosphere and study the impact of GNSS data for tomography.

A new tomography (H-representation), using solutions similar then used to solve the hydrogen atom solution to Schrödinger equation, provides and efficient way to represent the solution space. With as little as n=5 (55 coefficients) we obtained a fit of 7 TECU under the LEO track, using only LEO data, while with the addition of IGS ground data gave a fit of about 13 TECU. Using more coefficients, and adding smoothing constraints, the solutions gets become more accurate.

As expected, the addition of ground data from a few stations provided a better global fit. On the other hand, GNSS-R data improves significantly the ED results on the Orbital Ring (where they provide data) so it can be concluded that the addition of GNSS-R data can cover a crucial gap over the oceans, where "ground" (vertical) data is not available.

# 6 REFERENCES

(See also the Starlab Library at http://starlab.es)

[Arfken, GB and Weber HJ.] Mathematical Methods for Physicists. 1995. Academic Press.

[Komjathy 1996] Komjathy, A., and R. Langley, Improvement of a Global Ionospheric Model to Provide Ionospheric Range Error Corrections for Single-frequency GPS Users, presented at the ION 52nd Annual Meeting, Cambridge MA 19-21 June 1996.

[PIM] PIM 1.7 User guide, 1998, http://www.cpi.com/products/pim/ .

[Ruffini 2002b] Ruffini,G., Marco, J., Ruffini, L., Ionospheric (H-atom) Tomography: a Feasibility Study using GNSS Reflections, Slides presented at the Space Weather Workshop held at ESA/ESTEC, Dec 2002. Available at http://starlab.es.

[Ruffini 2002] Ruffini, G., WP3500 Paris Beta, Atmospheric Effects. Paris Beta ESA contract, 2002.

[Ruffini 2001] Ruffini, G., et al, Using GNSS reflections for Ionospheric Studies (zipped ps), *Space Weather Workshop*, ESA-ESTEC, Dec 2001

[Ruffini 1999] Ruffini, G., Cucurull, L., Flores, A., Rius, A., A PIM-Aided Kalman Filter for GPS Tomography of the ionospheric eletron content, Phys. Chem. Earth (C), Vol 24, No. 4, pp. 365-369, 1999.

[Ruffini 1998] Ruffini, G., A. Flores, and A. Rius, GPS Tomography of the Ionospheric Electron Content with a Correlation Functional, IEEE Transactions on Geoscience and Remote Sensing, vol 36, n. 1, January 19